\documentstyle[12pt]{article}
\begin{document}
\vspace{-2.0cm}
\begin{flushright}
IFUP-TH 7/98, SNBC/98-02-01 \\
hep-th/9802050\\
\end{flushright}
\bigskip
\begin{center}
{\large \bf ABELIANIZATION OF SU(N) GAUGE THEORY  WITH GAUGE INVARIANT DYNAMICAL VARIABLES ~\\
 AND MAGNETIC MONOPOLES
\footnote{Partially supported by EC Contract ERBFMRXCT97-0122 and by MURST}}
\end{center}

\bigskip

\begin{center}
\large {Adriano Di Giacomo\footnote{Electronic address:
digiacomo@pi.infn.it}
and Manu Mathur\footnote {Electronic address: manu@boson.bose.res.in, 
manu@ibmth.difi.unipi.it}} \\
\bigskip
${}^{2}$ Dipartimento di Fisica dell' Universit\^a and I.N.F.N, \\
Piazza Torricelli 2 Pisa, 56100 Italy \\
\bigskip
${}^{3}$ S. N. Bose National
Centre for Basic Sciences,  \\
JD Block, Sector 3, Salt Lake City, Calcutta  700091, India.\\  
\end{center}

\bigskip
\begin{center}
{\bf ABSTRACT}\\
\end{center}

\bigskip

\noindent It is shown that SU(N)  gauge theory coupled to adjoint Higgs 
can be explicitly re-written in terms of SU(N) gauge invariant dynamical 
variables with local physical interactions. 
The resultant theory has a novel compact abelian $U(1)^{(N - 1)}$ gauge 
invariance. The above abelian gauge invariance is related to the 
adjoint Higgs field and not to the gauge group SU(N). In this abelianized 
version  the magnetic monopoles carrying the magnetic charges of $(N - 1)$ 
types have a natural origin and therefore appear explicitly in the partition 
function as Dirac monopoles along with their strings. The gauge invariant 
electric and magnetic charges with respect to $U(1)^{(N-1)}$  gauge groups 
are shown to be vectors in root and co-root lattices  of SU(N) respectively. 
Therefore,  the Dirac quantization condition corresponds to SU(N) Cartan matrix 
elements being integers. We also study the effect of the $\theta$ term  in the 
abelian version of the theory. 
\\ 

\newpage 

\begin{center} 
{\bf I. ~~~INTRODUCTION}\\ 
\end{center}

The  subject of magnetic monopoles has been fascinating ever since they 
were proposed by Dirac \cite{dirac1} in 1931  to explain electric charge 
quantization in abelian theory. He showed that the quantization condition 
between electric (e) and magnetic (g) charges:

\begin{eqnarray} 
e g =  2 \pi n, ~~~~ n=0,\pm 1, \pm 2, ..... 
\label{cq} 
\end{eqnarray}  

\noindent  was both  necessary and sufficient  to quantize  an abelian 
theory in the presence of magnetic monopoles. As a consequence of these 
external magnetic charges, the abelian gauge group 
becomes compact. On the other hand, in the non-abelian SU(N) gauge theories 
the gauge group is defined to be compact independent of any charges. Like 
the abelian case above, the compactness of the non-abelian gauge group is again intimately 
related to the presence of magnetic monopoles. However, these magnetic charges are not visible  
in the original non-abelian Lagrangian but occur as topological excitations of the theory. 

Perhaps the most important role of the magnetic monopoles in physics 
is their  expected  role in the mechanism of quark confinement.  It is widely 
believed  that their condensation could provide an interesting and  sufficient 
framework to explain quark confinement along the lines of dual Meissner 
effect \cite{hooft2}. This conjucture has been tested explicitly 
in the context of much simpler compact lattice quantum electrodynamics 
(CLQED) where the magnetic degrees of freedom manifestly appear in the partition 
function. In the confining phase of CLQED, these magnetic charges condense in the 
vacuum leading  to a linear potential between the electric charges. 
It is generally hoped that the confinement of the color charges in the non-abelian theories 
will have its roots manifest in its abelianised version. If so, the qualitative picture of confinement in the 
non-abelian theories could be very similar to the confining phase of CLQED. 
Therefore, an important problem before studying the vacuum properties of the non-abelian theories,  
is to abelianize them so as to make the contribution of the topological magnetic degrees of 
freedom to the partition function explicit.  
A similar but simpler problem arises in the context of abelian Higgs theories with Abrikosov-Nielsen-Olesen (ANO) 
vortices as topological excitations. This problem has been widely studied in 
the past. In this abelian framework, the radial decomposition of the complex 
Higgs field has been extremely useful in extracting out the contribution of 
the ANO vortices from the partition function \cite{ano}. 
However, to our knoweldge, a  similar construction for the non-abelian gauge 
theories and its relevence to topological magnetic monopoles and hence to 
confinement is still lacking inspite of large amount of  literature on this subject. 
This will be the motivation and the subject of this paper. 
In the non-abelian SU(N) gauge theories, based on the suggestion 
of 't Hooft \cite{hooft1}, extensive work has been done in the past to 
study the topological magnetic monopoles in effective abelian theories   
via ``abelian projections". By abelian projection it is meant that the 
SU(N) gauge group is gauge fixed such 
that only its maximal Cartan subgroup $U(1)^{N-1}$ is left untouched. In 
these approaches with pure gauge theories  certain collective excitations 
of the theory act as an effective ``SU(N) adjoint Higg field" and 
the space time points where two of the eigenvalues are equal are the 
possible locations of the magnetic charges. However, unlike the abelian Higgs model 
or the compact lattice quantum electrodynamics, these issues have been 
largely  addressed at the level of kinematics and there 
has been very little progress at the level of dynamics besides Monte Carlo 
simulations \cite{kron}.   
In this paper we formulate the  idea of radial decomposition in the context of 
SU(N) gauge theory and study its consequences to the magnetic monopoles. A very 
qualitative discussion of its relevance to confinement is given at the end. 
For the sake of simplicity, we take the adjoint Higgs mentioned above not as composite 
field but as one of the microscopic field present in the Lagrangian. However, most 
of the results presented below are general. They depend only upon the adjoint transformation 
property of the Higgs  and not   on the detailed form of the  Lagrangian except on its gauge 
invariance.  Also, the low energy physics of the pure SU(N) gauge theory can be thought of as 
the large mass limit of the Higgs in the present formulation. We find that the non-abelian radial 
decomposition to be described below has  many resemblences with the 
corresponding abelian formulation along with some novel results. They are are summarised below: 

\begin{enumerate} 
\item  SU(N) gauge theories with adjoint Higgs can be  rewritten completely in terms of explicitly gauge 
invariant fields. This is similar to the abelian Higgs model (see section II), 

\item In terms of the above dynamical variables  novel {\it compact} abelian gauge invariance $U(1)^{N-1}$ 
({\it not a subgroup of SU(N)}) naturally emerges, 

\item As a consequence of abelianiasation,  the magnetic monopoles\footnote{These monopoles include both SU(N) 
t' Hooft Polyakov types which are  solutions of the classical equations of motions in the Higgs phase as well 
as the ones which are not the solutions 
and occur independent of the phases. The latter can be looked upon as local defects in space time and  might 
also contribute to the dynamics 
and the spectrum of the theory, e.g in the case of pure Q.C.D their condensation can lead to confinement of the 
colored gluons. The classical equations of motion or their solutions will not be used anywhere in this paper.} 
in the theory now are point like and their contribution to the partition 
function can be explicitly extracted. As expected,  this contribution is similar to the one proposed by 
Dirac in the context of abelian gauge theory with external magnetic charges \cite{dirac}. Moeover, they 
carry charges of  $(N-1)$ types  corresponding to the $(N -1)$ U(1) gauge groups and 
couple to their  corresponding abelian gauge fields with coupling proportional to $({1 \over e})$. 
This feature is again similar to the abelian Higgs model where the radial decomposition makes the contribution of the 
ANO  vortices to the partition function explicit and they appear with inverse coupling\cite{ano}.   

\item The Dirac quantization condition  has simple interpretation in terms of the geometrical properties of the  
roots and co-roots of the SU(N) group and corresponds to the scalar product of any root and co-root 
vectors being integers. 

\end{enumerate} 

The results and the techniques presented below are  independent of the space time dimension (d) but 
we illustrate the idea in d=4. The organisation of the article is as follows.  The section II 
is devoted to the kinematical aspects of the SU(N) adjoint Higgs and is independent of any non-abelian 
gauge invariance of a theory. 
It starts with the description of the basic idea of constructing SU(N) adjoint Higgs dynamics in the 
''body fixed frame" (BFF) 
along with its novelties. This  idea is partly borrowed from the rigid body classical dynamics 
where the angular motion of the rigid body is described in terms of the angular velocities of the BFF. 
After implementing the  radial decomposition of Higgs field,  
we will find the angular velocity description  useful in constructing the SU(N) gauge invariant variables.  
The section IIB is devoted to describe the inbuilt $U(1)^{(N-1)}$ abelian gauge invariance in the 
above frame. It is emphasized that this local abelian invariance is solely due 
to the description of the adjoint Higgs in terms of its angular velocities in the BFF. 
Hence, the above abelian invariance should not be confused with any subgroups of the initial SU(N) gauge group. 
Partly for this reason, we have introduced the SU(N) gauge group and the corresponding gauge fields only in the 
section III after describing the $U(1)^{(N-1)}$ abelian gauge invariance in detail. We must 
emphasize here that in the full gauge theory in the unitary  gauge with the Higgs oriented in a particular 
fixed direction, we recover all the results of t' Hooft \cite{hooft1}. In this sense what follows can be looked 
upon as the explicit gauge invariant formulation of t' Hoofts ideas at the level of partition function. In what 
follows, we will also be borrowing heavily from the work of t' Hoofts.  
In section II.C we describe the geometrical constraints on the 
angular velocities of the body fixed frame of any SU(N) adjoint Higgs field. Singularities in these 
constraints will eventually lead to topological magnetic charges in the theory. One important assumption through 
out this paper is that the singularities in the constraints in the section II.C occur at discrete space time 
points. The section II.D involves the computation 
of the Jacobians on going to the angular co-ordinates and then to the angular velocity descriptions. 
In section III we include the gauge fields. In this section, exploiting the SU(N) gauge transformation 
properties of the Higgs angular velocities and of the gauge fields we  construct the SU(N) gauge invariant 
and  $U(1)^{(N-1)}$ gauge covariant fields $Z^{\pm}_{\mu}(x)$ along with SU(N) gauge invariant $(N-1)$ photons 
of $U(1)^{(N-1)}$ gauge group. All the features discussed above are general and model independent. 
Only at the end of the section III we will introduce the SU(N) Higgs model with an arbitrary gauge invariant 
potential to implement these ideas explicitly at the level of the partition function. 
Therefore, these techniques  can also be applied to pure SU(N) Yang Mills theory where there is no 
microscopic Higgs field present in the Lagrangian. The role of Higgs is now played by a particular choice of  
composite gluonic field.  The section IV is devoted to study of the magnetic monopoles 
which are necessarily the locations where the the descripition of the Higgs field in terms of its angular velocities
fails.  At the end we summarize our results and discuss $\it{qualitatively}$ a probable relevance of the radial 
decomposition to the problem of confinement.

\newpage 

\begin{center}
{\bf II. ~~~ THE BODY FIXED FRAME, ANGULAR VELOCITIES  AND THE $U(1)^{(N - 1)}$ 
GAUGE INVARIANCE} 
\end{center} 

\vspace {0.5cm} 

We first describe our notations and some definitions which will be useful to describe 
SU(N)  gauge theory and the corresponding magnetic monopoles within a general framework.  
A  hermitian traceless $N \times N$  matrix $\phi$ is described by 
$(N^{2} -1)$ real parameters. The set of all such matrices will form the real vector 
space $R^{N^{2}-1}$. Any hermitian traceless matrix $\phi$ will be called 
a vector in $R^{N^{2}-1}$. In a particular basis,  $\phi$ will also be  
denoted by $\vec{\phi}$ where each of its $N^{2}-1$ elements being the components in 
that basis. The scalar product of two vectors $\phi_{1}$ and $\phi_{2}$ is 
defined by 

\begin{eqnarray}
(\phi_{1},\phi_{2}) \equiv  {1 \over 2} Tr (\phi_{1}\phi_{2}). 
\label{sp} 
\end{eqnarray} 

\noindent Here $Tr$ stands for the trace. The norm of a vector $\phi$ is $(\phi,\phi)$. 
The space $R^{(N^2 - 1)}$  can be spanned by any set $(N^{2} - 1)$ orthonormal (with respect 
to the scalar product $\left(\ref{sp})\right)$  basis vectors.  

In the Cartan Weyl basis they are constructed  such that 
$H_i, \left(i =1,2.....,(N - 1)\right)$ are the set of commuting elements and $E_{\pm \alpha}, 
\alpha = 1,......,{(N^{2} - N) \over 2}$ are the non-commuting ladder operators
In the standard notations \cite{wy}  their algebra is: 

\vspace{0.3cm}

\begin{eqnarray}
\left[H_{i},H_{j}\right]   & = &  0, \nonumber \\ 
\left[H_{i}, E_{\alpha}\right] &  = &  K_{i}(\alpha) E_{\alpha} \nonumber \\ 
\left[E_{\alpha},E_{\beta}\right]  & = & 
\vec{K}(\alpha).\vec{H} 
~~~ if ~~~ \alpha = - \beta \nonumber \\ 
                                    & =  &   N_{\alpha,\beta}^{\gamma} E_{\gamma} ~~~ otherwise.  
\label{cartan} 
\end{eqnarray} 

\vspace{0.3cm}

In (\ref{cartan}) the    $(N - 1)$ dimensional root correspoding to the ladder operator $E_{\alpha}$ is 
denoted by $\vec{K}(\alpha)$. $N_{\alpha,\beta}^{\gamma}$ are constants depending upon the group 
and are non-zero only if $\vec{K}(\alpha) + \vec{K}(\beta) = \vec{K}(\gamma)$. We will denote the Cartan subspace 
by H and the $N-1$ simple roots spanning H  will be denoted by $\vec{K}(\alpha_{s}),~ s=1,2,...,N-1$.  
The elements of the algebra (\ref{cartan}) are normalised such that $(H_{i},H_{j}) = \delta_{ij}, 
(E_{\alpha},E_{\beta}) = {(1 / 2) } \delta_{\alpha + \beta,0}$.  
From now onwards this basis will be called the {\it{``Space fixed frame"}} (SFF) because it is given 
in terms  of a set of constant (space time independent) elements.  
A convenient realisation of this basis, to be used often hereafter, is defined in terms of the $N^2$ matrices $e_{ab}$, 
(a,b =1,2.....,N) with matrix elements $\left(e_{ab}\right)_{cd} \equiv \delta_{ac} \delta_{bd}$.  In terms of these 
matrices 

\begin{eqnarray} 
\label{basis} 
H_{i} = A_{i} \left[\sum_{j=1}^{i} e_{jj} - i e_{i+1i+1}\right],~~~  \\   \nonumber 
E_{+\alpha} =  e_{ab},~~~~~~~ E_{-\alpha} =  e_{ba} ~~~ a < b.  
\end{eqnarray} 

\noindent In the representation (\ref{basis}) $A_{i} \equiv \sqrt{2 \over i(i+1)}$ are the normalisation 
constants and the root vectors are labelled by $\vec{K}_{ab}$. 
Another representation to be used is  the SU(N)  Hermitian Gell-Mann $\lambda$ matrices: 
$\lambda_{i} \equiv H_{i}, \lambda_{+\alpha} \equiv E_{+\alpha}+E_{-\alpha}$ and  $\lambda_{-\alpha} \equiv 
i(E_{+\alpha}-E_{-\alpha})$ satisfying $(\lambda_{a},\lambda_{b}) = \delta_{ab},  (a=1,2.......N^2 - 1)$\footnote 
{In what follows the indices h,i,j.. will take values from 1 to $N-1$ and the Greek indices $\alpha,\beta$ 
will vary from 1 to ${N^2 - N \over 2}$. The indeces a,b,c.. will vary from 1 to N as well as from 1 to 
$N^2 - 1$ which  will be clear from the context or will be explicitly mentioned.}.

Given a vector $\phi$,  we will now  define a  $(N^2-1)$ dimensional orthonormal 
``body fixed frame" (BFF) as follows: ~
We first note that  $\phi(x)$ being a hermitian traceless matrix, it
can be  unitarily transformed into the $(N - 1)$ dimensional Cartan sub-basis:  

\begin{eqnarray}
U(x) \phi(x) U^{-1}(x) & \equiv & \sum_{h=1}^{N - 1} \rho_{(h)} \lambda_{h}.  
\label{diag}
\end{eqnarray}

\noindent Here U(x) is a SU(N) matrix in the coset space ${SU(N) \over {U(1)}^{N-1}}$. Further 
the diagonalisation procedure (\ref{diag}) is unique upto the ${N!}$ arrangements of the N 
eigenvalues $(v_{1},v_{2}....,v_{N})$ of $\phi(x)$ along the diagonal. This permutation of 
the N eigenvalues form a permutation group of order $N!$ which is isomorphic to the 
the Weyl reflection group \cite{wy}  of the SU(N). Explicitly, the operator which interchanges 
the  the eigenvalue $v_{a}$ with $v_{b}$ (\ref{basis}) is

\begin{eqnarray} 
\omega_{[a,b]} = expi {\pi \over 2} \left( E_{+\alpha} +  E_{-\alpha}\right). 
\label{exch} 
\end{eqnarray} 
    
\noindent The exchange operator $\omega_{[a,b]}$  in (\ref{exch})  also corresponds to the reflction 
across the hyperplane perpedicular to the root $\vec{K}(\alpha)$. Each of the $N!$ choices 
of arranging the eigenvalues $v_a$ in  (\ref{diag}) correspond to choosing the set $\rho_{h}$ 
in one of the $N!$ Weyl chambers one for each element of the Weyl group.
A Weyl chamber ${\cal C}_{\omega}$ corresponding to the Weyl group element $w$ is 
defined through the $(N-1)$ simple roots as follows: 

\begin{eqnarray} 
{\cal C}_{\omega} = \left(h \in H,~ (\omega~h~\omega^{-1},K(\alpha_s)) > 0, s=1,2,...,N-1\right). 
\label{weyl}
\end{eqnarray} 
 
\noindent ${\cal C}_{\omega=1}$ is called the fundamental Weyl chamber (FWC).  The hyperplanes seperating the Weyl 
chambers are called the Weyl walls. A vector with two of the eigenvalues 
equal\footnote{In SU(N) gauge theories with $\phi(x)$ as the adjoint Higgs, this is a necessary condition for 
the magnetic monopoles \cite{hooft1}.} lies on one of the Weyl walls.  
Using  these  discrete Weyl reflections in U(x), it is always possible to choose $\rho_{(h)} \ge 0$.
In the basis (\ref{basis}),  the choice  with all the eigenvalues arranged in the decreasing order of magnitude 
from top to bottom corresponds to $\rho_{h} \ge 0, \forall h$. This can be seen by  the relation:

\begin{eqnarray}  
v_{a} - v_{b} = \vec{\rho}.\vec{K}_{a,b}
\label{order} 
\end{eqnarray} 

\noindent for $b = a+1, (a=1,2,....N-1)$ and using the $\vec{K}$ in the basis $(\ref{basis})$. This particular Weyl 
chamber with $\rho_{h} \ge 0, \forall h$ is characterised by the Weyl group element $\omega$:   

\begin{eqnarray} 
\label{pwc} 
\omega & = &  \omega_{[N,2]}\omega_{[N-1,3]}..........\omega_{[{N \over 2}+2,{N \over 2}]},~~~~~~~~~~~~~ N: even ~~~~\\ 
       & = &  \omega_{[N,2]}\omega_{[N-1,3]}..........\omega_{[{(N-1) \over 2}+2,{(N-1) \over 2}+1]},~~ N: odd \nonumber  
\end{eqnarray}

\noindent In the following the chamber (\ref{pwc}) will be called the positive Weyl chamber (PWC). In the case 
of SU(2) $\omega$ in (\ref{pwc}) is an unit operator and FWC is also the PWC: a positive real line. 
However, for SU(N), $N > 2$, the  $(N-1)$ positive $\rho_{i}$ are not completely independent. The eqn. 
$(\ref{order})$  implies the following $N-1$ inequalities defining the PWC:

\begin{eqnarray} 
\rho_{i} \ge \sqrt{{i-1 \over i+1}} \rho_{i-1} 
\label{ineq} 
\end{eqnarray}  
 
Having thus obtained the $(N-1)$ ``radii" for a Hermitian traceless matrix $\phi(x)$, we now  define a set of 
$(N^2 -1)$ orthonormal basis vectors $\xi^{a}(x)$ through the ``angular" degrees of freedom of $\phi(x)$:

\begin{eqnarray} 
\xi^{a}(x) \equiv U^{-1}(x)\lambda^{a} U(x).    
\label{bff} 
\end{eqnarray} 

\noindent  The $N^{2} - 1$ vectors in (\ref{bff})  form another orthonormal basis in $R^{N^{2} - 1}$ which is space 
time dependent.  The components of the BFF axis in the space fixed frame $(\lambda_{A})$ are given by $\xi^{a}_{A} 
=(\lambda_{A},\xi^{a})$ and are real. In the sequel, the components of an arbitrary vector ${v}$ along the 
$\xi^{\pm \alpha}$ 
will be denoted by ${v}^{\mp \alpha}$ and will be called the ``chiral components" of ${v}$.  
The  equation $(\ref{diag})$ defines $\phi(x)$ in terms of U and the radial variables $\rho_{i} 
\ge 0$ and is the non-abelian analogue of breaking a  complex abelian Higgs field into its radial and angular parts.  
All the topological properties of $\phi(x)$ are now contained in the unitary matrix U(x).   
As mentioned before the abelian analogue of $(\ref{diag})$  has been widely applied in abelian theories to study topological 
excitations:

\begin{enumerate} 

\item  In the case of 2-dimensional abelian Higgs model it  corresponds to writing the complex Higgs field 
$\phi(x) \equiv \rho(x) exp i \theta(x)$. If $A_{\mu}(x)$ is the U(1) gauge field and e is the electric charge 
then defining 
$\omega_{\mu}(x) \equiv \partial_{\mu} \theta(x)$, one can rewrite the theory 
in terms of the U(1) gauge invariant variables $e Z_{\mu}(x) \equiv A_{\mu}(x) - \omega_{\mu}(x)$ and the  
radial field 
$\rho(x)$.   The abelian field strength tensor in terms of $Z_{\mu}$ is $F_{\mu\nu}(Z) = \partial_{\mu} Z_{\nu}(x) 
- \partial_{\nu} Z_{\mu}(x) -{ 1 / e} (\partial_{\mu}\partial_{\nu} - \partial_{\nu}\partial_{\mu}) \theta^{sing}$. 
Here $\theta^{sing}$ is the multivalued part of the phase angle $\theta$ of $\phi$.  The last term in the field 
strength tensor above describes the ANO vortices. 

\item In 4 dimensions the radial decomposition of spin 0 magnetic charges $\phi(x) 
\equiv \rho(x) exp i \theta(x)$ was 
exploited to write down a manifestly Lorentz co-variant and local quantum field theory of these particles 
interacting with electric charges \cite{manu}. In this work once again the topological magnetic currents  
in the final theory were intimately related to the global angular behaviour of $\theta(x)$. 

\end{enumerate} 

Motivated by the above results in the context of abelian theories, we now formulate the idea of radial decomposition 
further in the context of non-abelian gauge theories and exploit it  to study the magnetic monopoles. The treatment 
will be very similar in spirit to that of the ANO vortices in the case of abelian Higgs model \cite{ano,emil}.  Its final 
consequences are already summarised in the section I. 

In the next section we will  express all the color vectors and their space time derivatives in  
the BFF basis. 
Being an orthonormal basis we know that it can only undergo rotations in space time. 
In other words, analogous to rigid body dynamics, these changes will be given 
in terms of the {\it angular velocities} of the BFF. By taking the space time derivative 
of (\ref{bff}) it is easy to see 

\begin{eqnarray}
\partial_{\mu} \xi^{a}(x) - i ~[\omega_{\mu}(x),\xi^{a}(x)] \equiv 
D_{\mu}\left(\omega(x)\right)_{ab}\xi^{b} = 0. 
\label{angvel} 
\end{eqnarray} 

In (\ref{angvel}), $\omega_{\mu}(x) \equiv i U^{-1}(x)\partial_{\mu}U(x)$ are the angular 
velocities of the BFF and   $D_{\mu}(\omega)$ are the ``covariant" derivatives with respect to the matter angular velocities. 
The meaning of covariant will be clear after we study the transformation properties of 
$\xi^{a}(x)$ and $\omega_{\mu}^{a}(x)$ under the abelian (section II.B) and the SU(N) gauge symmetries 
(section III). 

\vspace{0.8cm} 

\begin{center}
{\bf II.B ~~~ THE ABELIAN GAUGE INVARIANCE} 
\end{center} 

\vspace{0.5cm} 

The Higgs field in (\ref{diag}) depends only on $\xi^h$ (h=1,2,.....,$(N - 1)$) and the corresponding 
radial fields $\rho_{h}(x)$.  Therefore, the BFF basis defined in (\ref{bff}) is determined only upto 
local $U(1)^{(N - 1)}$ abelian gauge invariances corresponding to the rotations around each of $\xi^{h}(x)$. 
This implies that  any dynamics rewritten in the BFF will have  an inbuilt local $U(1)^{(N - 1)}$ invariance 
{\it independent of any gauge group}. This invariance is also compact.  Defining 
$U^{(H)}(x) \equiv U^{-1}(x) 
\left(exp~ i\sum_{h=1}^{N-1} ~\theta^{h}(x)\lambda^{h}\right) U(x)$, the  local abelian invariances are:    

\begin{eqnarray} 
\xi^{a}(x) & \rightarrow &  U^{(H)}(x) \xi^{a}(x) {U^{(H)}}^{-1}(x). 
\label{abinv}
\end{eqnarray} 

\noindent The computation of the transformation laws of the BFF (\ref{bff}) is straight 
forward. 
We denote  the $(N - 1)$ angles of the compact $U(1)^{N - 1}$ gauge group by $\vec{\theta}(x) \equiv 
\left(\theta^{1}(x),\theta^{2}(x),...\theta^{N - 1}(x)\right)$. 
The $U(1)^{(N - 1)}$ abelian gauge transformations  of the BFF basis vectors are given by:        

\begin{eqnarray} 
\xi^{h}(x)  &\rightarrow & \xi^{h}(x) \nonumber \\ 
\xi_{\pm \alpha}(x)  & \rightarrow & exp\left( - i~ \vec{K}_{\pm \alpha}.\vec{\theta}(x)\right)~~ \xi_{\pm \alpha}(x).  
\label{abinv2} 
\end{eqnarray} 

\noindent The vectors $\vec{K}_{\pm \alpha} \equiv \left(K^{1}_{\pm \alpha},K^{2}_{\pm \alpha},..........
,K^{N - 1}_{\pm \alpha}\right)$  are the root vectors defined in  (\ref{cartan}). 
The abelian transformations (\ref{abinv2}) induce the following transformations on the 
Cartan and the chiral components of the angular velocities:

\begin{eqnarray} 
\omega_{\mu}^{h}(x) & \rightarrow &  \omega_{\mu}^{h}(x) +  \partial_{\mu} \theta^{h}(x) \nonumber \\ 
\omega_{\mu}^{\pm \alpha}(x) &  \rightarrow  & exp\left( i~ \vec{K}_{\pm \alpha}.\vec{\theta}(x)\right) ~~
\omega_{\mu}^{\pm \alpha}(x). 
\label{angveltransab} 
\end{eqnarray} 

Note that the BFF  components of the angular velocities in the Cartan subspace transform like abelian 
gauge fields while its chiral components transform like matter fields with charges proportional to 
the corresponding root vectors.  Once again we emphasize that the abelian symmetries (\ref{abinv2}) and 
(\ref{angveltransab}) must not be taken as the subgroups of the gauge group SU(N) to be introduced 
later in the section III. Infact, any Higgs model, where the Higgs field is a vector in the internal space,
rewritten in the BFF in terms of its angular velocities    
will have (\ref{abinv2}) and (\ref{angveltransab}) as its local invariance.  A simple example is  the $\sigma$ 
model with global O(3) invariance. This model in its angular velocity description has a local U(1) invariance \cite{lus}.

\vspace{0.8cm} 

\begin{center}
{\bf II.C ~~~ THE GEOMETRICAL CONSTRAINTS} 
\end{center} 

\vspace{0.5cm} 

In the section III,  we would like to describe the dynamics of the adjoint Higgs  vector not by its 
orientation U(x) in  (\ref{diag}) but by its angular velocities. 
This, as will be shown in the section III will enable us to construct  SU(N) gauge dynamics 
in terms of explicit gauge invariant variables. However, the angular velocity description introduces 
extra variables. To begin with, we had the Higgs field with  the $N-1$ radial 
fields and the $N^{2}-N$ co-ordinates of the diagonalisation matrix U in the coset space $SU(N) / U(1)^{N-1}$ 
describing the orientation of the BFF with respect to the SFF. However, 
the angular velocities $\omega_{\mu}^{a}(x)$ have $4 \times (N^2-1)$ degrees of freedom.  Therefore, in terms of them  
the dynamics is highly constrained. Naively, these constraints on the angular velocities  
can be easily computed  through its defining equation (\ref{angvel}): 

\begin{eqnarray}
[D_{\mu}(\omega),D_{\nu}(\omega)] \xi^{a}(x) \equiv 0 ~~~~~ =>~~~~~  F_{\mu\nu}^{a}(\omega) = 0. 
\label{cons1} 
\end{eqnarray} 

In deriving the constraint $F_{\mu\nu}^{a}(\omega) = 0$ in (\ref{cons1}) we have assumed that ~ 
$(\partial_{\mu}\partial_{\nu} - \partial_{\nu}\partial_{\mu}) \xi^{a}(x) \equiv 0$. This  is not true as 
the BFF basis vectors in general can be functions of  multivalued fields in space time. 
Given a generic Higgs vector we would like to locate  the space time points where the right hand side of 
the constraint (\ref{cons1}) is different from zero. Towards this end, we characterize the 
diagonalization matrix U(x) by a set of $(N^2-1)$  parameters  denoted by 
$\vec{\Theta}(x)$.  The defining equation of $\omega_{\mu}(x)$ in (\ref{angvel})  now can be 
rewritten as 

\begin{eqnarray}  
\omega_{\mu}^a = - H^{a}_{b}(\vec{\Theta}(x))  \partial_{\mu} \Theta^{b}(x). 
\label{angvel3} 
\end{eqnarray} 

\noindent The eqn. (\ref{angvel3}) expresses the SFF components of the angular velocities in terms of the  matrix 
$H^{a}_{b}(\vec{\Theta}(x))$ which is defined through the composition 
functions $F[\vec{\Theta}_{1};\vec{\Theta}_2]$,  $\left(i.e ~~ U\left(\vec{\Theta}_{1}\right) 
U\left(\vec{\Theta}_{2}\right)~ \right) \equiv U\left(F[\vec{\Theta}_{1};\vec{\Theta}_2] \right)$  
of the SU(N) group \cite{sud} as follows: 

\begin{eqnarray} 
M^{a}_{b}(\vec{\Theta}(x))  \equiv   { \partial F^{a}[\vec{\Phi};\vec{\Theta}(x)] \over \partial\Phi^{b}}
|_{Lim \vec{\Phi} \rightarrow 0}  \nonumber \\ 
M^{a}_{b}(\vec{\Theta}(x)) H^{b}_{c}(\vec{\Theta}(x))  \equiv   \delta^{a}_{c}. ~~~~~~~~~~
\label{inver} 
\end{eqnarray} 

The equation (\ref{angvel3})  defining  the angular velocities is a straight forward generalization 
of the angular velocities defined in the case of rigid body classical dynamics \cite{sud} 
to field theory.  Now computing the field strength tensor for the angular velocities we 
find that

\begin{eqnarray} 
{\cal F}_{\mu\nu}^{a}\left(\omega\right) &  = & \left[H^{a}_{b,c}(\vec{\Theta}(x)) - H^{a}_{c,b}(\vec{\Theta}(x))+ f^{aef} H^{e}_{c}
(\vec{\Theta})H^{f}_{b}(\vec{\Theta})\right]\partial_{\mu}\Theta^{c}(x)  
\partial_{\nu} \Theta^{b}(x)  \nonumber \\
& + &  H^{a}_{b}(\vec{\Theta}(x)) \left(\partial_{\mu} \partial_{\nu} - \partial_{\nu} \partial_{\mu}\right) \Theta^{b}(x).
\label{ds1} 
\end{eqnarray}  

\noindent In (\ref{ds1}) $H^{a}_{b,c}(\vec{\Theta})$ is the partial derivative of $H^{a}_{b}(\vec{\Theta})$ with respect 
to ${\Theta}^{c}$.  However, because of the associativity property of the group, the first term in the square bracket 
is  zero. This term being zero is just   the  integrability condition of the partial differential equation satisfied 
by the composition function $F[\Theta_1;\Theta_2]$ due to associativity property of the group. Note that different 
co-ordinate systems chosen to describe the unitary 
matrix  will have their own set of structure constants and composition functions but they will all satisfy the above 
integrability condition.  For more details the reader is referred to \cite{sud}.  The necessary condition 
for the right hand side of (\ref{ds1}) being not equal to zero at some space time point $x_{0}$ is that atleast one of the 
parameters describing the unitary matrix U  should be multivalued function of space time. 
The sufficient condition is that the support $H^{A}_{B}(x_{0})  \neq 0$. 
In what follows, we will assume that the  above singular points  $x_{0}$ are discretely  and not continuosly 
distributed in the space time.
To incorporate these multivalued fields in the theory it is convenient to devide
$\Theta(x) \equiv \Theta^{[r]}(x) + \Theta^{[s]}(x)$.
Here $\Theta^{[r]}(x)$ and $\Theta^{[s]}(x)$ are single and multivalued
functions of space time and are defined by: 

\begin{eqnarray}
\oint_{{\cal{C}}\in\Delta{\Sigma_{\mu\nu}(x^{0}_{\mu})}}
\partial_{\mu} \Theta^{[s]}(x) dx_{\mu} & = & 2 \pi {\cal Z}, ~~ {\cal Z}: Integers  \nonumber \\
\oint_{{\cal{C}}\in\Delta{\Sigma_{\mu\nu}(x_{\mu})}}
\partial_{\mu} \Theta^{[r]}(x) dx_{\mu} & \equiv &  0 ~~  \forall x_{\mu} \in R^{4}. 
\label{defects}
\end{eqnarray}

\noindent Here ${\cal{C}}$ is a curve enclosing any surface
$\Delta{\Sigma_{\mu\nu}(x^{0}_{\mu})}$ around a singular space time point
$x^{0}_{\mu}$. 
So finally we see that a careful computation of the field strength tensor of the matter angular velocities 
gives\footnote{This computation also shows that the unitary transformation defined in (\ref{diag}) must not 
be confused with the gauge transformation (see section III) and trivially absorbed in the gauge invariant measure.}:

\begin{eqnarray}
{\cal F}^{a,(np)}_{\mu\nu}\left(\omega(x)\right) = H^{a}_{b}\left(\vec{\Theta}(x)\right) 
\left(\partial_{\mu}\partial_{\nu} - \partial_{\nu}\partial_{\mu}\right) \Theta^{[s]b}(x).  
\label{ds2} 
\end{eqnarray}  

\noindent We have attached an extra superscript (np) on ${\cal F}_{\mu\nu}^{a}(\omega)$ now and it stands for 
{\it ``non-perturbative"}.  Its origin will be clear when we discuss the full gauge theory.  
In section IV we will explicitly study the singular nature of (\ref{ds2}) and its consequences to 
the complete partition function of the gauge theory and in particular to magnetic monopoles.

\vspace{0.8cm} 

\begin{center}
{\bf II.D   ~~~ THE MEASURE} 
\end{center} 

\vspace{0.5cm} 

Before including the gauge fields, it is instructive to compute the Jacobian on going 
from the Euclidean description of the Higgs to the radial and the angular velocity description. 
After including the gauge fields in the section III and IV we will see that through  
this Jacobian we are led to $(N-1)$ types of Dirac terms describing the magnetic monopoles in the final abelian 
version of the SU(N) theory.  In the first step of this section we will 
compute the Jacobian from  the Eucledian $(\phi)$ to the angular and the radial co-ordinates defined in (\ref{diag}). In the 
second step we will further transform the angular part of this measure into the measure 
in terms of the angular velocities. Some of the results in the first part of this section are already in the 
literature \cite{shab}. However, for the sake of completeness and also because our interpretation of some of the 
terms in the final measure is different (see footnote 8),  we will reproduce them briefly below. 

The metric tensor in the Euclidean co-ordinates of the adjoint Higgs is defined by $(d\phi(x),d\phi(x))
= \delta^{ab}\phi_{a}\phi_{b}$. To compute the measure it is convinient to characterize the diagonalisation 
matrix $U(\Theta)$ in terms of the variables in the Cartan subalgebra H and the coset space ${G \over H}$: 

\begin{center} 
$U\left(\Theta(\theta,z)\right)  \equiv U_{H}(\theta) U_{{G \over H}}(z)$. 
\end{center} 

Here $[\theta_{i}(x), i=1,2...,N-1]$ and $[z_{\alpha}(x), 
\alpha =1,2,....N^{2}-N]$ are the variables characterizing the subgroup $U(1)^{N-1}$  and the coset 
space $SU(N) / U(1)^{N-1}$.  Defining $d\omega(z) \equiv i U_{{G \over H}}(z)^{-1} dU_{{G \over H}}(z) = 
- \lambda^{\alpha}(z)H_{\alpha\beta}(z)\delta z^{\beta}(x) - \lambda^{i}(z) \tilde{H}_{i \beta}(z)
\delta z^{\beta}(x)$,  $\left[\rho_{i}\lambda^{i}(x),\lambda^{\alpha}(x)\right] \equiv 
\omega_{\alpha\beta}(\rho)\lambda^{\beta}$ with $\omega(\rho)$ being a linear matrix in the radial 
variable $\rho(x)$.  Now  using  (\ref{diag}) a straightforward calculation leads to \cite{shab}:

\begin{eqnarray} 
(d\phi,d\phi)  =  d\rho_{i}d\rho_{j} \left(\lambda^{i},\lambda^{j}\right)   
+ \left(\left[\rho_{i}\lambda^{i},d\omega\right],\left[\rho_{i}\lambda^{i},d\omega\right]\right) \nonumber \\ 
 =  \sum_{i=1}^{N-1}  d\rho_{i}^{2} + \delta z^{\alpha}\left(H^{T}(z) \omega(\rho) \omega(\rho) H(z)\right)_{\alpha\beta} 
\delta z^{\beta}.~~~~~~~  
\label{xyz} 
\end{eqnarray} 

\noindent In (\ref{xyz}) $H^{T}(z)$ stands for the transpose of the matrix H defined by (\ref{angvel3}) in the coset 
space. 
Thus, the metric $g_{ab}$ in the new co-ordinates is block diagonal: a unit matrix in the Cartan 
subspace and is determined by the composition functions and the structure constants in the  coset space. 
The corresponding 
Jacobian $J(\rho,z)$ on going to angular variables is $|H(z)|~|\omega(\rho)|$ ($|A| \equiv$ det A). In the 
Cartan basis \cite{shab},  ~  $|\omega(\rho)| = \prod_{\alpha =1}^{{N^{2}-N} \over 2} (\vec{\rho}.\vec{K}(\alpha))^{2}$, 
where the product over $\alpha$ runs only over the positive roots. Therefore, 
the Jacobian $J(\rho,z)$ splits into the following radial and angular parts: 

\begin{eqnarray}  
 J(\rho,z) \equiv J(\rho)J(z); ~~~~~ 
J(\rho) = \prod_{\alpha =1}^{{N^{2}-N} \over 2} (\vec{\rho}.\vec{K}(\alpha))^{2}, ~~~~~ J(z) =  |H(z)|. 
\label{jacobian} 
\end{eqnarray} 

\noindent In (\ref{jacobian}) $\vec{K}(\alpha)$  are the positive roots. Therefore,  we finally get\footnote{In \cite{shab} 
the integration over the radial variables $\rho_{h}(x)$ is restricted to the fundamental Weyl chamber of SU(N)
and therefore has non-trivial boundaries. Moreover, the integration over z in (\ref{meas}) is treated as trivial 
and ignored taking it as the gauge group volume in the full adjoint Higgs gauge theory.}: 

\begin{eqnarray} 
\prod_{a=1}^{N^{2}-1} \int d\phi^{a}(x) = \Big[\prod_{i=1}^{N-1}\int^{\prime} d\rho^{i}(x)\Big] 
\prod_{\alpha=1}^{N^{2}-N \over 2} \left(\vec{\rho}.\vec{K}(\alpha)\right)^{2}\Big[\prod_{\alpha=1}^{N^{2}-N} 
\int dz^{\alpha}\Big] |H(z)|.  
\label{meas} 
\end{eqnarray}

\noindent In the first set of integrations in (\ref{meas}) $(\prime)$ is used to denote the 
contrained range of the radial integrations due to (\ref{ineq}). 
We would like to convert  the last two factors of the measure in (\ref{meas}) into the SU(N) invariant 
Haar measure over $\Theta(x)$, i.e.,   

\begin{center} 
$ \int {\cal D} U(\Theta) \equiv \prod_{a=1}^{N^{2} -1} \int d\Theta^{a} |H(\Theta)| 
= \prod_{i=1}^{N-1} \int d\theta^{i} \prod_{\alpha=1}^{N^{2}-N} dz^{\alpha} |H(\Theta(\theta,z))|$. 
\end{center}

%
%
%
%
%
%
This can be seen by exploting the $U(1)^{N-1}$ gauge invariance of the starting Euclidean measure. We make the 
$U(1)^{N-1}$ gauge transformation $z \rightarrow \Theta(\theta,z)$ and introduce an unity of the form,  
$\prod_{i=1}^{N-1} \left({1 / 2 \pi} \int d\theta_{i}\right) ~~ \equiv ~~ 1$ in (\ref{meas}). 
Thus we get the manifest SU(N) invariant measure: 

\vspace{0.3cm}

\begin{eqnarray} 
\prod_{a=1}^{N^{2}-1} \int d\phi^{a}(x) = \Big[\prod_{i=1}^{N-1}\int^{\prime} d\rho^{i}(x)\Big] 
\prod_{\alpha=1}^{N^{2}-N \over 2} \left(\vec{\rho}.\vec{K}(\alpha)\right)^{2} \int {\cal D} U(\Theta). 
\end{eqnarray} 
  
\vspace{0.3cm} 

In the second step, to go to the angular velocity description, we 
introduce an unity  in  the form of a $\delta$ function over the angular velocities and then integrate over the 
true gauge degrees of freedom $\Theta^{[r]}(x)$: 

\vspace{0.3cm} 

\begin{eqnarray}
\label{constraint} 
\int d\vec{\omega}_{\mu}^{a}(x) \delta\left(\vec{\omega}_{\mu}^{a}(x) -
\left(\xi^{(a)}, U(\Theta) \partial_{\mu} U^{-1}(\Theta)\right)\right)  \equiv  1  ~~~~~~~~~~~~~ \\ 
\int {\cal D} U(\Theta) \delta \left(\omega_{\mu}^{a} - \left( \xi^{a}, U(\Theta) \partial_{\mu}
U^{-1}(\Theta \right)\right) 
  =   \int {\cal D}\Theta^{[s]}\delta 
\left(F_{\mu\nu}^{a}\left(\vec{\omega}\right) 
+ {\cal F}_{\mu\nu}^{a{(\it{np})}}(\Theta^{[s]})\right).   \nonumber 
\end{eqnarray}

\vspace{0.3cm} 

For the sake of simplicity we have omitted to indicate the products over space time points, Lorentz 
and SU(N) group indeces in (\ref{constraint}).  The first equation in (\ref{constraint}) is just the 
definition of the angular velocities. The right hand side of the second equation is motivated by the SU(N) 
invariance of the Haar measure of both left and right hand side. 
Thus, in terms of the radial fields and the angular velocities: 

\vspace{0.3cm} 

\begin{eqnarray} 
\label{ang} 
\prod_{a=1}^{N^{2}-1} \int d\phi^{a}(x) & & = ~~~~  \Big[\prod_{i=1}^{N-1}\int^{\prime} d\rho^{i}(x) 
\prod_{\alpha=1}^{N^{2}-N \over 2} \left(\vec{\rho}.\vec{K}(\alpha)\right)^{2}\Big] \\ 
& & \int d\omega^{a}_{\mu}(x)  \int {\cal D}\Theta^{[s]} ~  
\delta
\left(F_{\mu\nu}^{a}\left(\vec{\omega}\right)
+ {\cal F}_{\mu\nu}^{a{(\it{np})}}(\Theta^{[s]})\right). \nonumber 
\end{eqnarray} 

\vspace{0.3cm}

We will see the relevance of (\ref{ang}) later  after including the gauge fields in the section III and in the 
context of the magnetic monopoles in the section IV. 

\vspace{0.8cm} 

\begin{center}
{\bf III.  ~~~ THE SU(N) GAUGE GROUP} 
\end{center} 

\vspace{0.5cm} 

Till now our discussion has been independent of any gauge group and the corresponding gauge fields. At this stage we 
intoduce the SU(N) gauge group acting on the adjoint Higgs $\phi(x)$ and gluonic fields $W_{\mu}(x)$ as: 

\begin{eqnarray} 
\phi(x) &  \rightarrow  & G(x) \phi(x) G^{-1}(x) \nonumber \\  
W_{\mu}(x) &  \rightarrow  &  G(x) W_{\mu}(x) G^{-1}(x) + i G(x) \partial_{\mu} G^{-1}(x). 
\label{gt1} 
\end{eqnarray}

\vspace{0.3cm}

Before introducing any particular model, it is useful to study the SU(N) gauge 
transformation properties  of the new variables $\xi^{a}$ and $\omega_{\mu}(x)$ we had introduced in the 
Section II.  From the gauge   transformation  of the adjoint Higgs and the equation (\ref{diag}) it is clear that 
all $\xi^{h}$ also transform like adjoint fields. If G(x) is the gauge transformation matrix in the fundamental 
representation of SU(N) then the orthonormality of the BFF basis vectors forces us to define:   

\vspace{0.3cm}

\begin{eqnarray} 
\xi^{a}(x)  & \rightarrow & G(x) \xi^{a}(x) G^{-1}(x)  \nonumber \\
\omega_{\mu}(x) &  \rightarrow & G(x) \omega_{\mu}(x) G^{-1}(x) + i G(x) \partial_{\mu} G^{-1}(x).
\label{abc} 
\end{eqnarray} 

\vspace{0.3cm}

It is easy to see from (\ref{abinv2}) that under $U(1)^{(N - 1)}$ local gauge transformations the 
components of $W_{\mu}(x)$ in the BFF 
$(W^{a}_{\mu}(x) \equiv (W_{\mu},\xi^{a})$ undergo the following induced transformations: 

\begin{eqnarray}
W^{h}_{\mu}(x) & \rightarrow & W^{h}_{\mu}(x),   \nonumber  \\
W^{\pm \alpha}_{\mu}(x) &  \rightarrow &  exp \left(i ~ \vec{K}^{\pm \alpha}.\vec{\theta}(x)\right) W^{\pm \alpha}_{\mu}(x). 
\label{efg}   
\end{eqnarray}

\noindent We now define the covariant gauge fields by 

\begin{eqnarray}
e Z_{\mu}(x) \equiv \omega_{\mu}(x) - W_{\mu}(x). 
\label{zz}
\end{eqnarray} 

In (\ref{zz}) we have explicitly introduced a constant e for later convenience. It will eventually be identified 
with the SU(N) coupling constant. As the BFF basis vectors $\xi^{a}(x)$ and $Z_{\mu}(x)$ both transform covariantly 
under SU(N) gauge transformation (see (\ref{gt1}), ({\ref{abc}) and (\ref{zz})),  the components of the vector 
$Z_{\mu}(x)$ in the BFF $Z_{\mu}^{a}(x) (\equiv (Z_{\mu},\xi^{a}(x))$ are explicitly SU(N) gauge 
invariant.  Therefore, any specific SU(N) gauge model with adjoint Higgs (which may be a composite 
field) rewritten in terms of $Z_{\mu}^{a}(x)$ variables will be explicitly gauge invariant.  
On the other hand, under $\left(U(1)\right)^{(N-1)}$ gauge transformations (see (\ref{angveltransab}) and 
(\ref{efg})):

\begin{eqnarray} 
Z_{\mu}^{\pm \alpha}(x) & \rightarrow & exp\left(i \vec{K}_{\pm \alpha}.\vec{\theta}\right)Z_{\mu}^{\pm \alpha}(x) 
 \nonumber \\
Z_{\mu}^{h}(x) & \rightarrow & Z_{\mu}^{h}(x) + {1 \over e} \partial_{\mu} \theta^{h}(x). 
\label{abtr} 
\end{eqnarray} 

\vspace{0.3cm} 
 
We notice that under $\left(U(1)\right)^{(N-1)}$ the $(N-1)$ components of $Z_{\mu}$ within the 
Cartan subalgebra of the BFF transform like abelian gauge fields. Following the language of t' Hooft, 
they will be called ``photons" and will be denoted by $A_{\mu}^{h}(x) (\equiv Z_{\mu}^{h}(x))$.
They are of  $N-1$ types and explicitly SU(N) gauge invariant. Further, 
the SU(N) gauge invariant chiral components 
$Z_{\mu}^{\pm \alpha}(x)$  transform like abelian matter fields with charges proportional to their root 
vectors. These gauge invariant electric charges we compactly denote by  
a set of ${N^2-N}$ vectors $ \vec{Q}_{[\pm \alpha]}$, each of them with $N-1$ components:  

\begin{eqnarray} 
\vec{Q}_{[\pm \alpha]} = e \vec{K}(\pm \alpha). 
\label{elch} 
\end{eqnarray} 

\noindent We can now define their abelian co-variant derivatives (to be used later) by 

\begin{eqnarray}
D_{\mu}(A) Z_{\nu}^{\pm \alpha}(x)  \equiv  \left(\partial_{\mu} 
-  \vec{A}_{\mu}(x).\vec{Q}_{[\pm \alpha]}\right) Z_{\nu}^{\pm \alpha}(x). 
\label{cov}
\end{eqnarray} 

\vspace{0.3cm} 

Having described the general idea and the framework we now introduce the SU(N) Higgs model and explicitly 
show its abelianization and as a consequence the emergence of $(N - 1)$ types of magnetic monopoles in the 
partition function. The Lagrangian is: 

\vspace{0.3cm}

\begin{eqnarray} 
{\cal L} =  - {1 \over 4 e^2} (F_{\mu\nu}(W),F_{\mu\nu}(W)) - {1 \over 2} (D_{\mu}\phi(x),D_{\mu}\phi(x)) + V(\phi(x)).   
\label{lagg}
\end{eqnarray} 

\vspace{0.3cm} 

In (\ref{lagg}) $\phi$ and $W_{\mu}$ are the hermitian traceless matrices describing the Higgs and the gluon 
fields respectively. $D_{\mu} \phi \equiv \partial_{\mu} \phi(x) - 
i ~[W_{\mu},\phi]$ and 
$F_{\mu\nu}(W) \equiv  \partial_{\mu} W_{\nu} - \partial_{\nu} W_{\mu} - i ~[W_{\mu},W_{\nu}]$. In (\ref{lagg}) e 
is the SU(N) coupling constant introduced in (\ref{zz}).  The Higgs potential can be any SU(N) gauge invariant 
function of the Higgs field and will 
not play any crucial role in what follows. 

At this stage we would like to trade off the Higgs field by its angular velocities and re-express (\ref{lagg}) completely 
in terms of the $\omega_{\mu}(x)$ and the gluonic fields $W_{\mu}(x)$. Using the $\delta$ functions in the measure (\ref{constraint}) 
over the matter angular velocities 
,  it is easy to see that the matter and the field strength tensor parts of the Lagrangian are given by:  

\begin{eqnarray} 
(D_{\mu}(W)\phi(x),D_{\mu}(W)\phi(x))  &  = &   
  e^{2}  \sum_{\alpha=1}^{(N^{2}-N) \over 2}\left(\sum_{h=1}^{(N - 1)}\rho_{(h)} K^{(h)}_{+\alpha}\right)^2 
Z_{\mu}^{+\alpha} Z_{\mu}^{-\alpha} \nonumber \\    
 & +  & ~~~~~~~ \sum_{h=1}^{N - 1} \left(\partial_{\mu} \rho_{(h)}\right)^2. 
\label{higl} 
\end{eqnarray} 

\vspace{0.4cm} 

\begin{eqnarray} 
{\left(F,F\right) \over e^2} = 
\sum_{\alpha =1}^{(N^2-N) \over 2}\left(D_{\mu}(A)Z_{\nu}^{+\alpha}-D_{\nu}(A)Z_{\mu}^{+\alpha}  
- e\sum_{\gamma,\delta=1}^{(N^2-N) \over 2}N^{\alpha}_{\gamma,\delta}Z_{\mu}^{+\gamma}Z_{\nu}^{+\delta}
+{1 \over e} {\cal F}^{\alpha,(np)}_{\mu\nu}\right).h.c 
\nonumber \\  
+\sum_{h=1}^{(N-1)}\left(\partial_{\mu}A_{\nu}^{h}-\partial_{\nu}A_{\mu}^{h} - {i \over 2}  
\sum_{\alpha =1}^{(N^2-N) \over 2} Q_{[+\alpha]}^{h}\left(Z_{\mu}^{+\alpha}Z_{\nu}^{-\alpha} - 
Z_{\nu}^{+\alpha}Z_{\mu}^{-\alpha}\right) + {1 \over e} {\cal F}^{h,(np)}_{\mu\nu}\right)^{2}.~~~   
\label{fst} 
\end{eqnarray} 

\vspace{0.4cm}

\noindent In (\ref{fst}) $N^{\alpha}_{\gamma,\delta}$ are the constants defined in (\ref{cartan}) and 
are non-zero iff $\vec{K}(+\gamma) + \vec{K}(+\delta) = \vec{K}(+\alpha)$. 
The covariant derivative $D_{\mu}(A)$ are defined in (\ref{cov}). 
We would like to emphasize the following interesting features of the final lagrangian given by 
(\ref{higl},\ref{fst}): 

\begin{enumerate}  

\item  The Lagrangian (\ref{higl}) and (\ref{fst}) is not only expressed in terms of explicitly SU(N) gauge 
invariant variables  and local\footnote{We are ignoring the term ${\cal F}^{a,(np)}$ describing the magnetic monopoles 
with their unphysical (non-local and non-gauge invariant) Dirac strings. This will be discussed in the Section IV.}
but also manifestly invariant under the abelian gauge group $\left(U(1)\right)^{(N-1)}$,  i.e under the 
transformations (\ref{abtr}). Both these results were expected to begin with: 

\begin{enumerate} 

\item the $U(1)^{N-1}$ invariance 
as a consequence of describing the dynamics of the Higgs in the BFF in terms of its angular velocities, 

\item The explicit SU(N) gauge invariance due to the fact that the components of a vector in the BFF $\xi^{a}(x)$ 
are explicitly gauge invariant.  
  
\end{enumerate} 
\item  Note that in (\ref{fst}) both (e) and $({1 \over e})$ appear. Therefore, ${\cal F}^{a,(np)}$  
is a non-perturbative term. In the next section we will show that it describes  the $(N-1)$ magnetic monopole 
of the SU(N) theory. 

\end{enumerate}  

The final partition function is easy to compute. The action (\ref{higl}) and (\ref{fst}) has only implicit dependence 
on $\omega_{\mu}^{a}(x)$ through 
$Z_{\mu}^{a}(x)$. Therefore, trading off the integration over the gauge fields  $W_{\mu}^{a}(x)$  in terms of the 
explicitly gauge invariant fields $Z_{\mu}^{a}(x)$ in (\ref{zz}), the angular velocity integrations over 
their delta functions in (\ref{constraint}) give rise to unity. 
This leaves us with a complete  gauge invariant description of the SU(N) adjoint Higgs theory with non-perturbative 
topological degrees of freedom manifest at the quantum level. 
In the following sections, we will further discuss the topological aspects of the theory i.e, the  configurations 
having non-zero measure over $\Theta^{[s]}$ in (\ref{ang}) and the associated 
non-perturbative terms proportional to the inverse SU(N) coupling constant in  (\ref{fst}).

\vspace{1.0cm}

\begin{center}
{\bf IV. ~~~  THE MAGNETIC MONOPOLES} 
\end{center} 

\vspace{0.8cm}

Having  abelianized the SU(N) gauge  theory to the $U(1)^{(N - 1)}$ gauge group, it is expected that we will encounter 
magnetic monopoles carrying $(N - 1)$ types of topological magnetic charges corresponding to   
to each of the U(1) groups. The presence of the magnetic monopoles can already be seen  in the non-perturbative 
terms with couplings proportional to ${1 \over e}$ in (\ref{fst}).   
{\it Note that the starting non-abelian theory (\ref{lagg}) in terms of $W_{\mu}(x)$ and $\phi(x)$ 
had no such ``magnetic" term in the lagrangian}. In the final $U(1)^{(N - 1)}$ abelianised  
version, it is also expected that the contribution of these magnetic terms to the action will be  
similar to the one  proposed  by Dirac in the context of abelian gauge theory with external magnetic 
charges \cite{dirac}. We will proceed to show how these results emerge naturally in this section. 
We first note that the $(N-1)$ Noether currents of the $U(1)^{N-1}$ gauge symmetries are given 
by:

\vspace{0.3cm}

\begin{eqnarray} 
J_{\mu}^{i}|_{[i=1,2,...,N-1]} =  {i} \sum_{\alpha = 1}^{{N^{2}-N \over 2}} Q^{i}_{[\alpha]}
\Big[\left(Z_{\nu}^{-\alpha} D_{\mu}(A) Z_{\nu}^{+\alpha} -c.c\right) +  \nonumber \\
 e \sum_{\gamma,\delta=1}^{{N^{2}-N \over 2}} N^{\alpha}_{\gamma,\delta} 
\left(Z_{\mu}^{-\gamma}Z_{\nu}^{-\delta}Z_{\nu}^{+\alpha} -c.c\right) + 
2 \partial_{\nu} \left(Z_{\mu}^{+\alpha}Z_{\nu}^{-\alpha} - c.c\right)\Big].
\label{noether}
\end{eqnarray} 

\vspace{0.3cm}

\noindent We also define the topological currents: 

\vspace{0.3cm}

\begin{eqnarray} 
K_{\mu}^{i} \equiv {1 \over e}  \partial_{\nu} {\tilde{\cal F}}_{\mu\nu}^{i,(np)}. 
\label{top} 
\end{eqnarray} 

\vspace{0.3cm}
 
\noindent Now the $(N-1)$ abelian field strength tensors:

\vspace{0.3cm}

\begin{eqnarray} 
F_{\mu\nu}^{i}|_{[i=1,2,...,N-1]} = \partial_{\mu} A_{\nu}^{i} - \partial_{\nu} A_{\mu}^{i} - {1 \over e} 
{\cal F}_{\mu\nu}^{i,(np)}, 
\label{abfst} 
\end{eqnarray} 

\vspace{0.3cm}

\noindent satisfy  their corresponding ``Maxwells Equations" with topological magnetic sources, i.e: 

\begin{eqnarray} 
\partial_{\mu} F_{\mu\nu}^{i} =  J_{\nu}^{i}~~, ~~~~~~~~ \partial_{\mu} \tilde{F}_{\mu\nu}^{i} \equiv K_{\nu}^{i}, 
\label{maxwell} 
\end{eqnarray} 

\vspace{0.3cm}

\noindent In (\ref{top}) and (\ref{maxwell}) ${\tilde F}_{\mu\nu} \equiv {1 \over 2} \epsilon^{\mu\nu\rho\sigma} 
F_{\rho\sigma}$. 
The first set of the equations in (\ref{maxwell}) are the equations of motion of the $(N-1)$  $A_{\mu}(x)$ abelian 
gauge fields and the second set contains $(N-1)$ Bianchi identities. 
The question now is to identify the generic space time points where the topological magnetic 
currents $K_{\mu}^{i}$ do not vanish, i.e, the world lines of magnetic monopoles.  
These are the points where the two of the eigenvalues of $\phi(x)$ are equal and the 
radial decomposition (\ref{diag}) breaks down \cite{hooft1}. The eqn. (\ref{order})  
implies that at such points the Higgs field lies on one of the domain walls enclosing the PWC and 
the angular measure (\ref{ang}) vanishes.  In the neighbourhood of these singular points $(x_{0})$ 
the Higgs field lies in one of the $(N^{2} - N) / 2$ SU(2) subgroups of SU(N)  \cite{hooft1} and can always 
be written as: 

\vspace{0.3cm}

\begin{eqnarray} 
\phi_{[\alpha]}(x) \equiv  \phi(x_{0}) + |\epsilon(x)| \left[\phi_{[\alpha]}^{h} h_{[\alpha]} +  
\left(\phi_{[+ \alpha]} e_{[-\alpha]} + \phi_{[- \alpha]} e_{[+\alpha]}\right)\right]. 
\label{aa} 
\end{eqnarray}

\noindent  In (\ref{aa}) $\phi(x_{0})$ is a constant diagonal matrix which lies on one of the domain walls   
enclosing PWC,  $\epsilon(x)$ is a field $\rightarrow 0$ as $x \rightarrow x_{0}$ and  

\begin{eqnarray}
h_{[\alpha]} \equiv \sum_{i=1}^{N-1} {K^{i}(\alpha) H^{i} \over (\vec{K}(\alpha).\vec{K}(\alpha))}, ~ 
e_{[\pm \alpha]} \equiv  {1 \over \sqrt{\vec{K}(\alpha).\vec{K}(\alpha)} } E_{[\pm \alpha]}  
\nonumber \\   
\phi_{[\alpha]}^{h}(x) = cos {\tilde \theta}(x),~~~~~~ \phi_{[\pm \alpha]}(x) = sin {\tilde \theta}(x) 
exp \pm i {\tilde \psi}(x).   
\label{conf} 
\end{eqnarray} 

\noindent In (\ref{conf}) $\vec{K}(\alpha)$ is one of the $N-1$ roots $(\vec{K}_{1,2},\vec{K}_{2,3},....
\vec{K}_{N-1,N})$ enclosing the PWC and $\tilde\theta$ and $\tilde\psi$ are the polar and azimuthal angles 
in the SU(2) subspace corresponding to the root $\vec{K}(\alpha)$. Now the diagonalization matrix U in 
(\ref{diag}) can be written as: 

\vspace{0.3cm}

\begin{eqnarray}
U_{[\alpha]}(x) = exp\left(\left({e_{[+\alpha]} - e_{[-\alpha]} \over \sqrt{2}}\right) ~
{{\tilde\theta}(x) \over 2}\right)~~exp~ {i \over 2} \left( h_{[\alpha]} {{\tilde\psi}}(x)\right).
\label{eulan} 
\end{eqnarray}

\vspace{0.3cm}

\noindent The angular velocities for the above $(N -1)$ types of configurations in the BFF are given by: 

\vspace{0.3cm}

\begin{eqnarray} 
\label{mag} 
\omega_{\mu,[\alpha]}^{\beta} \approx  0 ~~~~~~~~~~~~~~~~~~~~~~~~~~~~~~~~~~~~~~~\\ 
\omega_{\mu,[\alpha]}^{h}|_{[h=1,2,...,N-1]}  =  -~ cos\left({\tilde\theta}\right)~ 
\left({2 K^{h}({\alpha}) \over 
\vec{K}(\alpha).\vec{K}(\alpha)}\right) \partial_{\mu} {\tilde \psi}^{sing} + regular~ terms.  \nonumber 
\end{eqnarray}  

\vspace{0.3cm}

\noindent In (\ref{mag}) ${\tilde \psi}^{sing}$  is the singular part of ${\tilde \psi}$ in (\ref{conf}) having 
non-zero support only along the polar axis  $(\tilde{\theta} = 0/\pi)$.  The symbol   $\approx 0$ means zero upto regular 
terms which do not contribute to ${\cal F}^{a,(np)}_{\mu\nu}$. 
From (\ref{mag}) the non-perturbative  terms describing the magnetic monopoles are 

\begin{eqnarray} 
\label{haha} 
{\cal F}_{\mu\nu}^{\alpha,(np)}(\omega_{[\alpha]})    & \approx 0 &   \\
{\cal F}_{\mu\nu}^{h,(np)}(\omega_{[\alpha]})|_{h=1,2,.....N-1} &   = &   - \hat{K}^{h}(\alpha) cos{\tilde \theta}(x)(\partial_{\mu} \partial_{\nu} 
- \partial_{\nu}\partial_{\mu}){\tilde\psi}^{sing}(x)  \nonumber \\  
&  =  &  \pm \hat{K}^{h}(\alpha) (\partial_{\mu} \partial_{\nu} - \partial_{\nu}\partial_{\mu}){\tilde{\psi}}^{sing}(x). \nonumber  
\end{eqnarray} 

Thus, these magnetic charges are proportional to the co-root vectors, $\vec{\hat{K}}(\alpha) \equiv \left({2 \vec{K}({\alpha}) / 
\vec{K}(\alpha).\vec{K}(\alpha)}\right)$. 
The last equation in (\ref{haha}) shows the presence of the Dirac strings along the internal polar axes, 
($\theta(x)=0,\pi$) if the azimuthal field  $\psi(x)$around it  has a non-trivial homotopy index ${\cal Z}$ (\ref{defects}).  
These  Dirac strings carry  singular fluxes  of amount $ ({2 \pi /e}) \vec{\hat{K}}(\alpha){\cal Z}$  towards the points where 
$\theta$ flips from 0 to $\pi$ (i.e the $\epsilon(x)$ in (\ref{aa}) vanishes and two of the eigenvalues are equal).   
These are the locations of the  Dirac monopoles \cite{hooft1}  with quantized magnetic 
charges in the units of ${4 \pi \over e} \vec{\hat{K}}(\alpha)$.  
Note that the Dirac strings are not gauge invariant  but the world lines of the magnetic monopoles 
are gauge invariant. The magnetic charge can now be expressed as a set of $N-1$ vectors, $\vec{M}_{[\pm \alpha]}$, 
each with $N-1$ components: 

\vspace{0.3cm}

\begin{eqnarray} 
\vec{M}_{[\pm \alpha]} = {4 \pi \over e} {\cal Z} \vec{\hat{K}}(\pm \alpha) 
\label{mach} 
\end{eqnarray} 

\vspace{0.3cm}

\noindent In (\ref{mach}) ${\cal Z}$ is the homotopy index of the mapping $S^{1}_{physical} \rightarrow S^{1}_{internal}$ 
provided by the multivalued $\psi^{sing}$ field. The equations (\ref{elch}) and (\ref{mach}) imply the Dirac quantization 
condition: 

\begin{eqnarray} 
\vec{Q}_{[\pm \alpha]}.\vec{M}_{[\pm \beta]}  = 4 \pi {\cal Z}_{[\pm \alpha,\pm \beta]}; ~~~~ \forall~ \alpha,\beta. 
\label{dqc} 
\end{eqnarray} 

\vspace{0.3cm}

\noindent Above ${\cal Z}_{[\pm \alpha,\pm \beta]}$  is the set of all integers and we have used the fact that the 
scalar products of the root and co-root vectors are integers. Expanding the root and co-root vectors in 
terms of simple roots and co-roots with integer coefficients, the Dirac quantization condition (\ref{dqc}) can be viewed 
as a simple consequence of the fact that the Cartan matrix elements are integers.  In \cite{eng} a quantization 
condition similar to (\ref{dqc}) was derived for the adjoint Higgs model of any compact simple Lie group but for a very 
special solution of the classical equations of motion. In \cite{eng} the following equations:

\begin{eqnarray} 
D_{\mu} F^{a \mu\nu}(W) =0, ~~ D_{\mu} \phi^{a}(x) = 0, ~~ {\partial V(\phi) \over \partial \phi^{a}} = 0 
\label{ps} 
\end{eqnarray} 

\noindent are considered. The solutions of (\ref{ps}) also satisfy the classical equations of motions. In the broken 
phase where the the SU(N) group is broken to its subgroup T, the equations (\ref{ps}) admit the monopole 
type solutions satisfying $\partial_{\nu}^{2} W_{\mu}^{a} =0$. The index a takes values in the subgroup $\bar{H}$ which is 
the Cartan subgroup of SU(N) conjugated by the unbroken group T. For such special solution,  invoking the Wu-Yang formulation 
the condition (\ref{dqc}) was obtained as the condition of single valuedness of the gauge transformations connecting 
the two vector potentials in the overlapping region.

In the case of SU(2), K=2 and $\hat{K} =1$ and therefore the magnetic monopoles are quantized in the units 
of $4 \pi / e$, i.e $g = {4 \pi / e} {\cal Z}$. 
Moreover,  as a special case, the famous SU(2) t' Hooft Polyakov hedgehog solitonic configurations \cite{hooft3} 
in the present abelian formulation 
corresponds to $K_{\mu}^{mag} = \left(\delta^{3}(\vec{x}),\vec{0}\right)$ in (\ref{top}) and hence 
to a point like Dirac monopole located at the origin where $\vec{\phi}(x) =0$ \cite{manu1}. We now 
discuss the full partition function in the special case of SU(2). In the case of SU(N) we have $N-1$ 
types of monopoles and the procedure trivially generalises except the for the  inequalities (\ref{ineq}). 
These inequalities can be accounted for by putting a step function in the radial measure.     
The integration over $\Theta^{sing}$ in (\ref{ang}) is the integration over the singular azimuthal 
angles $\psi(x)$. We will now rewrite it in terms of the Dirac strings.  
Let $X^{i}_{\mu}(\sigma_{1})$ be the gauge invariant world lines of the monopoles 
and $m^{i}$ a set of integers describing their magnetic charges in the units of 
${4\pi \over e}$. The  monopole current is  given by 
$K_{\mu}(x) = \sum_{i=1}^{\infty} m^{i} \int d\sigma_{1} {dX^{i}_{\mu}(\sigma_{1})
 \over d\sigma_{1}} \delta^{4}(x-X^{i}(\sigma_{1}))$. 
Therefore,  the magnetic term in (\ref{maxwell}) can be written as  

\begin{eqnarray} 
\tilde{{\cal F}}_{\mu\nu}^{a=3,(np)}(x) = \pm{4\pi \over e} \sum_{i=1}^{\infty} m^{i} \int 
d^{2}\sigma \epsilon^{\alpha\beta}\left(\partial_{\alpha}X^{i}_{\mu}(\sigma) 
\partial_{\beta}X^{i}_{\nu}(\sigma)\right) \delta^{4}\left(x-X^{i}(\vec{\sigma})\right). 
\label{dirr}
\end{eqnarray}      

\noindent  This is exactly  the term introduced  by Dirac \cite{dirac} in the 
context of abelian gauge theory where point particles carrying magnetic 
charges were put  by hand. 
Noticing that for SU(2) in (\ref{higl}) and (\ref{fst}) the constants $N_{\alpha,\beta}^{\gamma}$ 
in (\ref{cartan}) are zero, K=2, and Q=e, we find: 

\begin{eqnarray}
Z  & = & \sum_{m_{1},..,m_{\infty}} \prod_{i=1}^{\infty} \int dX_{\mu}^{i}
(\vec{\sigma}) J(X_{\mu}) \int \rho^{2} d\rho\int dZ_{\mu}^a   exp-S\left(
\rho,\vec{Z}_{\mu},X^{i}_{\mu}\right) \nonumber \\ 
S & = & \int\Big[{1 \over 4} \left(F_{\mu\nu}(A)\right)^{2} 
+ {1 \over 4}\left(D_{\mu}(A)Z_{\nu}^{+} - D_{\nu}(A)Z_{\mu}^{+}\right).h.c 
+  {ie \over 2}F_{\mu\nu}Z_{\mu}^{+}Z_{\nu}^{-} \nonumber \\
& - & {1 \over 16}\left(Z_{\mu}^{+} Z_{\nu}^{-} - h.c\right)^{2}   
+  {1 \over 2}\left(e^{2} \rho^{2}Z_{\mu}^{+}Z_{\mu}^{-}
 +   \left(\partial_{\mu}\rho\right)^{2}\right) + V(\rho)\Big] d^{4}x  
\label{final}
\end{eqnarray}

\noindent Here J(X) is the Jacobian \cite{emil} due to the change of the measure to 
the string world sheet.  The above partition function is manifestly invariant under 
U(1) gauge transformations. This is an exact result (with the assumption that the singular 
points where the Higgs field vanishes are discrete) and no 
gauge fixing has been done. In the broken 
phase where $\rho(x)$ = constant + fluctuations,  the 
above partition function describes the interaction of photon with the 
charged massive spin 1 gauge bosons and magnetic monopoles.  
The physical fields here  are explicitly gauge invariant 
and  have the right electric charges under $U(1)$. 
It is easy to see that in   the  unitary gauge  
$\phi(x) = \sigma_{3}$ which can be chosen only locally,  we recover the standard 
results.  

At this stage, having an exact abelian theory with magnetic monopoles 
in the partition function, we can also convert them into dyons 
by adding the CP violating and SU(N) gauge invariant $\theta$ term \cite{wit2} in the 
action (\ref{lagg}):

\vspace{0.3cm}

\begin{eqnarray} 
\Delta{\cal{L}}  = {e^{2} \over 32\pi^{2}} \theta  \sum_{i=1}^{N-1} \epsilon^{\mu\nu\rho\sigma}
F_{\mu\nu}^{i}(A_{\mu})F_{\rho\sigma}^{i}(A_{\mu}).  
\label{theta}
\end{eqnarray} 

\vspace{0.3cm}

\noindent Here $\theta$  is an angle in the range $[0,2\pi]$. 
The Maxwells equations (\ref{maxwell}) are now modified and acquires a $\theta$ dependent term: 
$\partial_{\mu}F_{\mu\nu}^{i}  =   J^{i}_{\nu} + 
{e^{2} \over 8 \pi^{2}} \theta K^{i}_{\nu}$.
Therefore, all the magnetic charges with strengths $\vec{M}_{[\alpha]}$ 
also acquire electric charges 
$\Delta \vec{Q}^{i}_{[\alpha]} = {e \over 2 \pi} \theta \vec{M}^{i}_{[\alpha]}$ 
leading to generalized  Schwinger quantization condition. Note that the  
$\theta$ term added above is not a surface term because of the Dirac strings. 

\vspace{0.8cm} 

\begin{center}
{\bf V. ~~~  DISCUSSION AND SUMMARY} 
\end{center} 

\vspace{0.5cm} 

Increasing evidence emerges from lattice simulations that dual superconductivity is the 
mechanism of color confinement, together with a better understanding of the nature 
of the monopoles involved \cite{adriano}. No similar results exist, for known reasons,
in the coninuum. Any better formulation of the theory can be helpful to progress in that 
direction. 
In this work we have shown that the SU(N)  non-abelian analogue of the radial decomposition 
widely used in the abelian Higgs  theories exists. We find it interesting that many features 
of the abelian theories are also hidden  in the non-abelian theories. 
Some of the outcomes like interpretation of the Dirac quantization condition in terms 
of elementary group theory results are already emphasised in the text. On the other 
hand from the point of view of color confinement,  probably the most 
interesting aspects are a) being able to make magnetic monopoles manifest in the partition 
function, b) the explicit non-abelian gauge invariance of all the fields in the partition 
function and emergence of the novel compact abelian invariances. The final rewritten partition function 
has $(N-1)$ photons and the corresponding 
electrically and magnetically charged fields, all explicitly color neutral. This 
was the picture proposed by t' Hooft \cite{hooft1} 
to find the macroscopic variables in Q.C.D.. However, his proposal of abelianization 
was at kinematical level through gauge fixing with the residual abelian 
gauge group as the subgroup of the initial gauge group.  

As mentioned in the introduction, 
all types of magnetic monopoles are defined through some particular choice of the effective adjoint Higgs 
field. All these choices are correlated.  Almost all the techniques presented here will go through for all of them. The difference 
being that the final partition function in (\ref{final}) will get suitably modified  
with more complicated form of (\ref{higl}) and the measure. However, the term  (\ref{fst})
which contains the  magnetic monopole interactions {\it will not change}.   
Thus, having a better control over the magnetic monopoles in the present framework,  
it is worthwhile to re-discuss {\it qualitatively} the  t' Hooft-Mandelstam \cite{hooft2} conjucture of 
confinement via dual Meissener effect.  We emphasise that most of the ideas  below are taken 
from \cite{hooft2,hooft4} and already well known. Our purpose here is  essentially to tie them 
with the present formulation and to the work done in \cite{manu}.  
For simplicity we restrict to the case of SU(2) and discuss only the issues related 
to the confining phase.   

The question of color confinement is  
the question  of $U(1)^{N-1}$ charge neutrality of the physical spectrum at large distances. 
The condensation of the magnetic monopoles is a sufficient framework to explain confinement 
vis dual Meissner effect. 
Before addressing this question of condensation in the present formulation, a  relevant 
problem is to eliminate the non-local Dirac strings at the level of dynamics. 
In other words to rewrite the final $U(1)^{N-1}$ gauge theory in a manifestly Lorentz co-variant and 
manifestly local  form.   Infact, this problem is very general.  It  will arise in any abelian theory with 
magnetic charges. This issue was resolved in \cite{manu} in the context of simpler single abelian theory of 
a spin 0 magnetically charged particle interacting 
with photon and other electrically charged  matter. In this work, roughly speaking,  
the Dirac strings were eliminated from the partition function by the use of an antisymmetric tensor 
field $H_{\mu\nu}(x)$.  
In the present context also, probably  the integration over the  Dirac strings 
in (\ref{final}) can be eventually re-written in terms of an integration over 
an anti-symmetric tensor field $H_{\mu\nu}$ with an appropriate Jacobian. 
This can be naively seen by  regulating the Dirac string  in (\ref{abfst}) on a lattice by  
spreading its flux  over a plaquette.  The field strength tensor in (\ref{abfst})  
now corresponds to\footnote{We are completely ignoring the Jacobian J(X) in (\ref{final}) 
for the qualitative purposes here.}:  

\begin{eqnarray} 
F_{\mu\nu}(n) = \Delta_{\mu} A_{\nu}(n) - \Delta_{\nu} A_{\mu}(n) - {4 \pi \over e} H_{\mu\nu}(n) 
\label{lat} 
\end{eqnarray} 

\noindent In (\ref{lat}) $H_{\mu\nu}(n)$ are the integer valued fields on a lattice, n are 
the lattice sites and $\Delta_{\mu}$ is the lattice difference operator. The naive continuum 
limit of (\ref{lat}) reproduces (\ref{abfst}). The  topological magnetic current  
in (\ref{top}) now corrsponds to:

\begin{eqnarray} 
K_{\mu}(n) = {\left(4 \pi \over e\right)} \Delta_{\nu} \tilde{H}_{\mu\nu}(n).  
\label{top2}
\end{eqnarray} 
 
\noindent This is also similar to the case of magnetic monopoles in compact quantum electrodynamics on a lattice 
with Wilson action in its Villain form \cite{pol1,ban}. As suggested by t' Hooft,    
one should eliminate all the electric charges, as accurately as possible,  by computing light-light-magnetic 
charge  scattering amplitudes and writing down the corresponding effective lagrangian for the photons with magnetic charges
\cite{hooft4}. The 
theory at this stage will be described in terms of the following microscopic degrees of freedoms: 

\begin{enumerate} 

\item The neutral  radial degrees of freedom $\rho(n)$ defined in (\ref{diag}) taking 
values from 0 to $\infty$,    

\item Anti-symmetric tensor fields $H_{\mu\nu}(n)$ describing the magnetic charges 
interacting with photons described by the magnetic vector potentials 
$A_{\mu}(n)$ (i.e, the magnetic field $\vec{B} \equiv \vec{\nabla} \times \vec{A}$).  

\end{enumerate} 

\noindent 
The monopole creation 
operators are still  non-local (\ref{top2}). Therefore, one should perform a 
duality transformation at this stage to get a local creation operator for the 
monopoles.  The duality transformation on the photon (in terms of the magnetic 
vector potential $A_{\mu}(x)$)  coupled to $H_{\mu\nu}(x)$ through its topological current and a  
neutral radial degree of freedom should eventually lead to a complex scalar field $\phi(x)$ minimally coupled 
to the photons (in terms of electric vector potential $\tilde{A}_{\mu}(x)$, i.e $\vec{E} \equiv 
\vec{\nabla} \times \vec{\tilde{A}}$)  through its Noether current.  
This picture was atleast true for a similar but simpler abelian theory \cite{manu}.  The Noether 
current above is the current associated with the ``dual" U(1) invariance of the  
$(\phi(n),\phi^{*},\tilde{A}_{\mu}(n))$ system. 
This final version of the theory will be an  effective Landau Ginzberg model of 
superconductivity in its dual form. 
The question of color confinement will be the question of the spontaneous symmetry breaking 
of the dual U(1) invariance. 
Note that for SU(N) both $\rho$ and $H_{\mu\nu}$ 
are $(N-1)$ in number and the dual description should lead to $(N-1)$ types of charged scalar 
fields. The inequalities (\ref{ineq}) for SU(N) with $N \ge 3$ make it distinct from SU(2). 
Its physical consequences, if any, are not clear to us.   

\vspace{0.5cm}
\hrule 
\vspace{0.5cm}
      
Acknowledgement: Manu Mathur acknowledges I.N.F.N. - Italy for the fellowship during 
which this work was done. It is also a pleasure for M.M to acknowledge Daniele Francesconi,  
Sara Ongaro and Alvise Santangelo for the pleasant time in Italy.

\end{document}